\documentclass{article}    

\usepackage{graphicx}

\title{\textbf{Auxiliary nRules of Quantum Mechanics}}  
\author{Richard Mould\footnote{Department of Physics and Astronomy, State University of New York, Stony Brook,
\mbox{New York} 11794-3800; http://ms.cc.sunysb.edu/\~{}rmould}}  
\date{}    
   
\begin{document}             

\maketitle              

\begin{abstract}

\end{abstract}    

Standard quantum mechanics makes use of four auxiliary rules that allow the Schr\"{o}dinger solutions to be related to laboratory
experience -- such as the Born rule that connects square modulus to probability.  These rules (here called the \emph{sRules}) lead
to some unacceptable results.  They do not allow the primary observer to be part of the system.  They do not allow individual
observations (as opposed to ensembles) to be part of the system.  They make a fundamental distinction between microscopic and
macroscopic things, and they are ambiguous in their description of secondary observers such as Schr\"{o}dinger's cat.

The \emph{nRules} are an alternative set of auxiliary rules that avoid the above difficulties.  In this paper we look at a wide
range of representative experiments showing that the nRules adequately relate the Schr\"{o}dinger solutions to empirical
experience.  This suggests that the sRules should be abandoned in favor of the more satisfactory nRules, or a third  auxiliary
rule-set called the \emph{oRules}.

\section*{Introduction}
Quantum mechanics traditionally places the observer `outside' of the system being studied, and refers only to ensembles of data. 
The theory does not refer to the primary data obtained in individual trials.  So standard quantum mechanics is incomplete for two
reasons -- it excludes the primary observer \emph{and} the primary data.  For a theory that claims to be a fundamental law of
nature, it excludes too much of nature.

Standard theory also makes a distinction between microscopic and macroscopic systems, and this is ambiguous at best.  In addition,
if one tries to use a quantum mechanical superposition of states to define an ontology in an individual trial, and if a secondary
observer like Schr\"{o}dinger's cat is included in the system, then the result will not only be paradoxical, it will be wrong.  The
cat is not in a dual state of consciousness at any time during that famous experiment.

Standard quantum mechanics posits four auxiliary rules that lead to these unacceptable  results.  These rules (called the
\emph{sRules}) are: (\textbf{1}) Quantum measurement occurs only when a quantum mechanical microscopic (or possibly mesoscopic)
system engages a macroscopic measuring device.  (\textbf{2}) Primary data is collected during individual trials of this kind in
which single eigenvalues are chosen by a stochastic process.  (\textbf{3}) This choice is followed by a state reduction (i.e., the
collapse) in which a stochastically chosen eigenvalue is the sole survivor.  And (\textbf{4}), the probability of choosing a given
eigenvalue is equal to its square modulus (i.e., the Born rule).  These auxiliary rules are not contained in the Schr\"{o}dinger
equation.  They are supplementary instructions that tell us how to use the Schr\"{o}dinger equation\footnote{These are Copenhagen
sRules.  There may be any number of variations.  Their essential feature is the Born Rule that establishes the \emph{only} link
between  observation and the symbols of the theory.}.

Other auxiliary rules are possible -- rules that avoid the difficulties described above.  There are at least two such rule-sets
called the \emph{nRules} and the \emph{oRules} that correctly relate Schr\"{o}dinger's equation to observation.  Both are
ontologically based, for they place the primary observer `inside' the system -- or at least they allow for that possibility.  With
either of these rule-sets (as in classical physics) the observer of an external system can extend the system to include himself so
he can become a continuous part of the wider system.  That is not possible with the auxiliary rules of standard quantum mechanics,
where the observer is only allowed to ``peek" at the system from time to time. 

In addition, the nRules provide an ontological description of individual trials, so they include the primary data obtained in each
trial.   They give us a running description of each trial.  They are sequentially deterministic except for the time interval between
stochastic choices, inasmuch as they cannot predict when the next stochastic choice will occur. The nRules are more deterministic
than the sRules, but they are less deterministic than classical physics.  The oRules are the least deterministic.  Both
the nRules and the oRules present an ontology for individual trials that removes the ambiguity that is now associated with
Schr\"{o}dinger's cat.

The proposed rule-sets do not make a fundamental distinction between microscopic and macroscopic things.  Each has four rules that
apply equally to all parts of nature.  As a result, both microscopic and macroscopic systems may be said to experience quantum
jumps; and in the nRule case, microscopic states can sometimes undergo state reduction.  Examples are given in this and related
papers.  Neither one of the proposed rule-sets directly includes the Born Rule that connects probability with square modulus; for in
both cases, probability is introduced only through \emph{probability current}.

This paper is concerned only with the nRules.  Their adequacy is demonstrated in a number of cases, and their properties (described
above) are made apparent.

\section*{Ontology and Epistemology}

The method of this paper differs from that of traditional quantum mechanics in that it sees the observer in an ontological rather
than an epistemological context.  Traditional or standard quantum theory (i.e., Copenhagen) places the observer outside of the
system where operators and/or operations are used to obtain information about the system.  This is the epistemological model shown
in \mbox{Fig.\ 1}.  

\begin{figure}[h]
\centering
\includegraphics[scale=0.8]{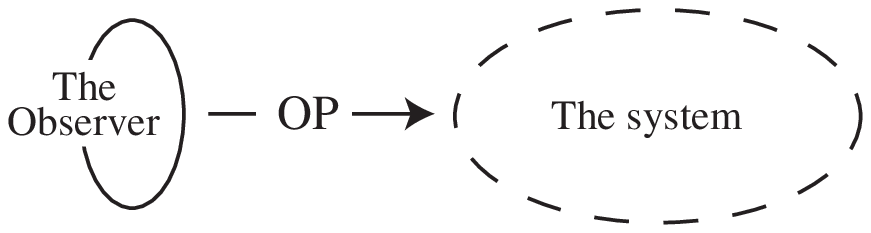}
\center{Figure 1: Epistemological Model (Copenhagen)}
\end{figure}

The large OP in Fig.\ 1 might be a mathematical `operator' or a corresponding physical `operation'.  The observer makes a
measurement by choosing a formal operator that is associated with a chosen laboratory operation.  As a result, the observer is
forever outside of the observed system -- making operational choices.  The observer is forced to act apart from the system as one who
poses theoretical and experimental questions to the system, and he can only get answers  through `instantaneous'
contacts with the system at a given time.  This model is both useful and epistemologically sound.  

	However, the special rules developed in this paper apply to the system by itself, independent of the possibility that an observer
may be inside, and disregarding everything on the outside.  This is the ontological model shown in \mbox{Fig.\ 2}.

\begin{figure}[h]
\centering
\includegraphics[scale=0.8]{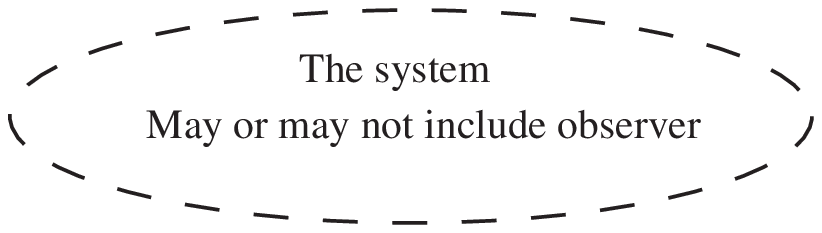}
\center{Figure 2: Ontological Model (requires special rules)}
\end{figure}

A measurement occurring inside this system is not represented by a formal operator.  Rather, it is represented by a measuring
device that is itself part of the system.  If the sub-system being measured is $S$ and a detector is $D$, then a measurement
interaction is given by $\Phi = SD$.  If an observer joins the system in order to look at the detector, then the system becomes
$\Phi = SDB$, where $B$ is the brain state of the observer.  Contact between the observer and the observed is continuous in this
case.

The ontological model is able to place the observer inside the universe of things and give a full account of his conscious
experience there.  It is a departure from traditional quantum mechanics and has three defining characteristics:  (1) It includes
observations given by $\Phi = SDB$ as described above, (2) it allows all conscious observations to be continuous, and (3) it rejects
the long-standing Born interpretation of quantum mechanics by introducing probability (only) through the notion of
\emph{probability current}. 

Quantum mechanical measurement is  said to refer to ensembles of observations but not to individual observations.  In this
paper I propose a set of four \emph{nRules (1-4)} that  apply to individual measurements in the ontological model.  I claim that
they are a consistent and complete set of rules that can give an ontological description of any individual measurement or
interaction in quantum mechanics.  These rules are not themselves a formal theory of measurement.  I make no attempt to understand
\emph{why} they work, but strive only to insure that they do work. Presumably, a formal theory can one day be found to explain these
rules in the same way that atomic theory now explains the empirically discovered rules of atomic spectra, or in the way that current
theories of measurement aspire to merge with standard quantum mechanics, or make the neurological connection with conscious
observation.

\section*{The oRules}
 Other papers \cite{RM1, RM2, RM3} propose another set of rules called the oRules (1-4).  These are similar to the nRules
except that the basis states of reduction are confined to observer brain states, reflecting the views of Wigner and von Neumann. 
Like the nRules, they introduce probability through the notion of `probability current' rather than through square modulus, and they
address the state reduction of conscious individuals in an ontological context, thereby giving us an alternative quantum friendly
ontology.  In Ref.\ 3 they are called simply the \emph{rules (1-4)}.  Like the nRules, the oRules are not a formal theory of
measurement  for they require a wider theoretical framework to be understood.  I do not finally choose one of the rule-sets or
propose an explanatory theory.  I am only concerned with how state reduction might occur in each case.  For the above stated
reasons, both of these rule-sets are more acceptable than the sRules; and so far as I am aware, there is no observation that can
distinguish between the two.

\section*{The Interaction: Particle and Detector}

		Before introducing the nRules, we will apply  Schr\"{o}dinger's equation to a `microscopic' particle interacting with a
`macroscopic' detector in order to see what difficulties arise.  These two objects are assumed to be initially independent and
given by the equation 
 
\begin{equation}
\Phi(t)=exp(-iHt)\psi_i\otimes d_i
\end{equation}
where $\psi_i$ is the initial particle state and $d_i$ is the initial detector state.  The particle is then allowed to pass over the
detector, where the two interact with a cross section that may or may not result in a capture.  After the interaction begins at a
time $t_0$, the state is an entanglement in which the particle variables and the detector variables are not separable.  

The first component of the resulting system is the detector $d_0$ in its ground state prior to capture, and the second, third, and
fourth components are the detector in various states of capture   given by $d_w$, $d_{m}$, and $d_d$.  
\begin{equation}
\Phi(t \ge t_0) =\psi(t)d_0 + d_w(t) \rightarrow d_{m}(t) \rightarrow d_d(t)
\end{equation}
where $d_w(t)$ represents the entire detector immediately after a capture when only the window side of the detector is affected, and
$d_d(t)$ represents the entire detector when the result of a capture has worked its way through to the display side of the
detector.   The middle state $d_{m}(t)$ represents the entire detector during stages in between, when the effects of the capture
have found their way into the interior of the detector, but not as far as the display. 

The state $\psi(t)$ is a free particle as a function of time, including all the incoming and scattered components.  It does no harm
and it is convenient to let $\psi(t)$ carry the total time dependence of the first component, and to let $d_0$ be normalized
throughout The first component in Eq.\ 2 is a superposition of all possible scattered waves of $\psi(t)$ in
product with all possible recoil states of the ground state detector, so $d_0$ is a spread of detector states including all the
recoil possibilities together with their correlated environments.  Subsequent components are also superpositions of this kind. 
They include all of the recoil components of the detector that have captured the particle. 
  
There is a clear discontinuity or ``quantum jump" between the two components $d_0$ and $d_w$ at the detector's window interface. 
This discontinuity is represented by a ``plus" sign and can only be bridged by a stochastic jump.  The remaining evolution from
$d_w(t)$ to $d_d(t)$ is connected by ``arrows"  and is continuous and classical.  These three detector states develop in time and
may be represented by the single component 
\begin{displaymath}
d_1(t) = d_w(t) \rightarrow d_{m}(t)\rightarrow d_d(t)
\end{displaymath}
 so
\begin{equation}
\Phi(t \ge t_0)=\psi(t)d_0 + d_1(t) 
\end{equation}
The capture state $d_1(t)$ in Eq.\ 3 is equal to zero at $t_0$ and increases with time\footnote{Each component in Eq.\ 3 has an
attached environmental term $E_0$ and $E_1$ that is not shown.  These are orthogonal to one another, insuring local decoherence. 
But even though Eq.\ 3 may be decoherent locally, we  assume that the macroscopic states $d_0$ and $d_1$ are fully coherent when
$E_0$ and $E_1$ are included.  So Eq.\ 3 and others like it in this paper are understood to be coherent when universally
considered.  We  call them ``superpositions", reflecting their global rather than their local
properties.}$^,$\footnote{Superpositions of macroscopic states have been found at low temperatures \cite{JRF}.  The components of
these states are locally coherent for a measurable period of time.}$^,$\footnote{Equation  3 can be written with coefficients
$c_0(t)$ and $c_1(t)$ giving $\Phi(t \ge t_0) = c_0(t)\psi(t)d_0 + c_1(t)d_1$, where the states $\psi(t)$, $d_0$, and
$d_1$ are normalized throughout.  We let $c_0(t)\psi(t)$ in this expression be equal to $\psi(t)$ in Eq.\ 3, and let $c_1(t)d_1$ be
equal to $d_1(t)$ in Eq.\ 3.}$^,$\footnote{It is important to realize that
the interaction Hamiltonian can only connect $\psi(t)d_0$ with the window state $d_w(t)$, which is the \emph{launch state} of the
activated detector.  It cannot pour probability current directly into a state that is more dynamically advanced, like $d_m(t)$ or
$d_d(t)$.  Therefore, at every moment during the current flow, a new state $d_w(t)$ is launched, since the state that was launched
immediately before that time has moved on (dynamically) to become $d_m(t)$. This means that the component $d_1(t)$ in Eq.\ 3 is  a
superposition of a continuum of all the detector states that have been launched at all previous times during the interaction.  For
the time being we will ignore this complication and come back to it (in Footnote 6) after we have examined this case from the point
of view of the nRules.}.

\section*{Add an Observer}
Assume that an observer is looking at the detector in Eq. 1 from the beginning.  
\begin{displaymath}
\Phi(t)=exp(-iHt)\psi_i\otimes D_iB_i
\end{displaymath}
where $B_i$ is the observer's initial brain state that is entangled with the detector $D_i$.  This brain is understood to
include \emph{only} higher order brain parts -- that is, the physiology of the brain that is directly associated with consciousness
after all image processing is complete.  All lower order physiology leading to $B_i$ is assumed to be part of the detector.  The
detector is now represented by a capital $D$, indicating that it includes the bare detector  \emph{plus} the low-level
physiology of the observer.

Following the interaction between the particle and the detector, we  have
\begin{eqnarray}
\Phi(t \ge t_0) &=& \psi(t)D_0B_0 + D_w(t)B_0 \rightarrow D_{m}(t)B_0 \rightarrow D_d(t)B_1 \hspace{.5cm}  \\
\mbox{or} \hspace{.5cm} \Phi(t \ge t_0) &=& \psi(t)D_0B_0 + D_1(t)B_1 \nonumber
\end{eqnarray}
where $B_0$ is the observer's brain when the detector is observed to be in its ground state $D_0$, and $B_1$ is the brain when the
detector is observed (only at the display end) to be in its capture state $D_1$.   As before, a discontinuous quantum jump is
represented by a plus sign, and the continuous evolution of a single component is represented by an arrow.  

If the interaction is long lived compared to the time it takes for the signal to travel through the detector in Eq.\ 4 (as in the
case of a long lived radioactive decay), then the superposition in that equation might exist for a long time before a capture.  
This means that there can be two active brain states of this observer in superposition, where one sees the detector in its ground
state and the other simultaneously sees the detector in its capture state.  Equation 4 therefore invites a paradoxical
interpretation like that associated with Schr\"{o}dinger's cat.  This ambiguity cannot be allowed.  It is not acceptable on
empirical grounds.  An observer who watches a detector in these circumstances will \emph{not} experience dual states of
consciousness.  

The nRules of this paper must not only provide for a stochastic trigger that gives rise to a state reduction, and describe that
reduction, they must also insure than an empirical ambiguity of this kind will not occur.

\section*{The nRules}
	The first rule establishes the existence of a stochastic trigger.  This is a property of the system that has nothing to do with
the kind of interaction taking place or its representation.  Apart from making a choice, the trigger by itself has no effect on
anything.  It initiates a state reduction only when it is combined with nRules 2 and 3.

\vspace{.4cm}

\noindent
\textbf{nRule (1)}: \emph{For any subsystem of n complete components in a system having a total square modulus equal to s, the
probability per unit time of a stochastic choice of one of those components at time t is given by $(\Sigma_nJ_n)/s$, where
the net probability current $J_n$ going into the $n^{th}$ component at that time is positive.}

\noindent
[\textbf{note}: A \emph{complete component} is a solution of Schr\"{o}dinger's equation that includes all of the (symmetrized)
objects in the universe.  It is made up of \emph{complete states} of those objects  including all their state variables.  If
$\psi(x_1, x_2)$ is a two particle system with inseparable variables $x_1$ and $x_2$, then $\psi$ is considered to be a single 
object.  All such  objects are included in a complete component.  A component that is a sum of
less than  the full range of a variable (such as a partial Fourier expansion) is not complete. This is why representation does not
matter to the stochastic trigger.]

\noindent
[\textbf{note}: Functions are not normalized in this treatment.  Instead, probability currents are normalized at each moment of
time by dividing $J$ by the value of $s$ at each moment of time. 

\vspace{.4cm}

The second rule specifies the conditions under which \emph{ready states} appear in solutions of Schr\"{o}dinger's equation.
These are understood to be the basis states of a state reduction. Ready states are always ``underlined" in this treatment.

\vspace{.4cm}

\noindent
\textbf{nRule (2)}: \emph{If a noncyclic interaction produces complete components that are discontinuous with the initial 
component, then all of the new states that appear in these components will be ready states.}

\noindent
[\textbf{note}: A cyclic interaction between two component is one that produces continuous oscillations.  A \emph{noncyclic
interaction} goes in one direction only.

\noindent
[\textbf{note}: Continuous means continuous in all variables. Although solutions to Schr\"{o}dinger's equation change
continuously in time, they can be \emph{discontinuous} in other variables -- e.g., the separation between the $n^{th}$ and the $(n +
1)^{th}$ orbit of an atom with no orbits in between.  A discontinuity  can also exist between macroscopic states that are locally
decoherent.   For instance, the displaced detector states $d_0$ (ground state) and $d_w$ (window capture state) appearing in
\mbox{Eq.\ 2} are discontinuous with respect to detector variables.  There is no state in between.  Like atomic orbits, these
detector states are a `quantum jump' apart.]

\noindent
[\textbf{note}: The \emph{initial component} is the first complete component that appears in a given solution of Schr\"{o}dinger's
equation.  A  solution is defined by a specific set of boundary conditions.  So Eqs.\ 1 and 3 are both included in the single
solution that contains the discontinuity between $d_0$ and $d_1$, where Eq.\ 1 (together with its complete environment) is the
initial state.  However, boundary conditions change with the collapse of the wave function.  The   component that
survives a collapse will be complete, and will be the initial component of the new solution.]

\noindent
[\textbf{note}: If a noncyclic interaction does not produce complete components that are discontinuous with the
initial component, then the Hamiltonian will develop the state in the usual way, independent of these rules.]

\vspace{.4cm}

The collapse of a wave function and the change of a ready state to a \emph{realized} state is provided for by nRule (3).  If a
complete state is not `ready' it will be called `realized'.  We therefore introduce dual state categories where ready
states are the basis states of a collapse.  They are on stand-by, ready to be stochastically chosen and converted by nRule (3) to
realized states.  In this paper, realized states are \emph{not} underlined.    

\vspace{.4cm}

\noindent
\textbf{nRule (3)}: \emph{If a component is stochastically chosen during an interaction, then all of the ready states that result
from that interaction (using nRule 2)  will become realized, and all other components in the superposition will be immediately
reduced to zero.}

\noindent
[\textbf{note}: The claim of an immediate (i.e., discontinuous) reduction is the simplest possible way to describe the collapse
of the state function.  A collapse is brought about by an instantaneous change in the boundary conditions of the
Schr\"{o}dinger equation, rather than by the introduction of a new `continuous' mechanism of some kind.]

\noindent
[\textbf{note}:  This collapse does not preserve normalization.  That does not alter probability of subsequent
reductions because of the way probability per unit time is defined in nRule (1); that is, current $J$ is divided by the total square
modulus.  Again, currents are normalized in this treatment -- not functions.]

\noindent
[\textbf{note}: If the stochastic trigger selects a component that does not contain ready states, then there will be no nRule (3) 
state reduction.]

\vspace{.4cm}

Only positive current going into a \emph{ready component} (i.e., a component containing ready states) is physically meaningful
because it represents positive probability.  A negative current (coming out of a ready component) is not physically meaningful and
is not allowed by nRule (4).  Without this restriction, probability current might flow in-and-out of one ready component and into
another.  The same probability current  would then be `used' and `reused'.  Given  the above rules, this would 
distort the total probability of a process.  If the nRules are to work, the total integrated positive current (divided by $s$) must
be no greater than 1.0.  To insure this we say     

\vspace{.4cm}

\noindent
\textbf{nRule (4)}: \emph{A ready component cannot transmit probability current.}

\vspace{.4cm}
Although it can receive current that increases its square modulus, a ready state is dynamically terminal.  It cannot develop beyond
itself.  If a ready state $\underline{S}_1(t)$ evolves from a state $S_0(t)$, then Schr\"{o}dinger's equation $H[S_0(t) +
\underline{S}_1(t)] = i\hbar\frac{d}{dt}[S_0(t) + \underline{S}_1(t)]$ will change $\underline{S}_1(t)$ in the usual way. 
However, nRule (4) will prevent the creation of a second order component.  The state $\underline{S}_1(t)$ is time dependent because
its square modulus increases \emph{and} because it reflects the dynamical changes  coming from $S_0(t)$ at every moment of time, 
but  it does not advance dynamically on its own.  It is the launch state of a new solution of Schr\"{o}dinger's equation if and when
it is stochastically chosen, and it will contain all the initial conditions of that solution.  Those initial conditions are not
applied until the moment of choice. 

While no theoretical reason can be given to explain Rule(4), its important consequences are demonstrated throughout the rest of
this paper.

\section*{Particle/Detector Revisited}
When the nRules are applied to the particle/detector interaction, Eq.\ 2 becomes 
\begin{displaymath}
\Phi(t \ge t_0) = \psi(t)d_0 + \underline{d}_w(t)
\end{displaymath}
where the quantum jump (+ sign)  is discontinuous and noncyclic, where \mbox{nRule (2)} requires that
$\underline{d}_w(t)$ is a ready state, and where the other components of the detector ($d_m$ and $d_d$) are zero because nRule (4)
will not allow $\underline{d}_w(t)$ to pass current to them.  At $t_0$, the first component is maximum and the second is zero, but
$\underline{d}_w(t)$ increases in time because of the probability current flow from $\psi(t)d_0$.  These components are
complete because the environment of each (not shown) is assumed to be present.

	The ready component in this equation is time dependent because of its increase in square modulus \emph{and} because it duplicates
the moment-to-moment changes that occur within the detector (such as molecular changes) \emph{although} it will not advance
dynamically to $d_m$.  So long a $\underline{d}_w(t)$ is a ready state, it will not pass current along to its successors.  As
indicated above, it is a function that contains the boundary conditions of the next solution of Schr\"{o}dinger's equation -- that
is, the `collapsed' solution that is realized when $\underline{d}_w(t)$ is stochastically chosen.  So the effect of nRule (4) is to
put the boundary conditions of the `next' solution on hold until there is a stochastic hit.  The nRules  then launch the new
solution with boundary conditions that are inherited at that moment from the old solution.  

	 	If there is a stochastic hit at time $t_{sc}$, then a continuous classical evolution will give
\begin{displaymath}
\Phi(t\ge t_{sc} > t_0)= d_w(t) \rightarrow d_m(t) \rightarrow d_d(t) 
\end{displaymath}
so the capture is eventually registered at the display end of the detector.

	 If there is \emph{no} stochastic hit on $\underline{d}_w(t)$ it will become a \emph{phantom} component.  A component is a phantom
when there is no longer probability current flowing into it (in this case because the interaction is complete), and when there can
be no current flowing out of it because it is a ready component that complies with nRule (4).  A phantom component can be dropped
out of the equation without consequence.  Doing so only changes the definition of the system -- it changes the total square modulus
$s$ that normalizes the current in nRule (1).  This is the same kind of redefinition that occurs in standard practice when one
chooses to renormalize a system at some new starting time.  Keeping a phantom is like keeping the initial system.  Because of nRule
(3), kept phantoms are reduced to zero whenever another component is stochastically chosen.  

	The nRules place this particle/detector system in an ontological setting, for they give an \emph{insider's} answer to the
probability question.  The nRules ask:  What is the probability that $\underline{d}_w(t)$ will be stochastically chosen during the
next time interval $dt$; and then, what is the probability that it will be stochastically chosen during the time interval $dt$
after that, etc?  The nRules are always concerned with what happens next.  Probabilities that are associated with the second order
states like $d_m(t)$ and $d_d(t)$ are ruled out by nRule (4) until the fate of $\underline{d}_w(t)$ is determined.  This is the
most important consequence of nRule (4) -- it does not allow a stochastic leap over the \emph{next} (ready) state of the system. 
The insider is concerned with the temporal ordering of states, and the nRules address that concern.

In contrast, the sRules ask:  What is the probability of finding $d_0$, $d_w$, $d_m$, or $d_d$ at some finite time $T$ after
$t_0$?  This is an outsider's question.  It is the question asked by one who can only observer the system at distinct and finitely
separated times like $t_0$ and $T$.

\section*{Particle/Detector/Observer Revisited}
To see how the nRules carry out a particle capture when an observer is a witness, we apply them to the first row of Eq.\ 4.  As
before, this only affects the first two components
\begin{equation}
\Phi(t \ge t_0)=\psi(t)D_0B_0 + \underline{D}_w(t)B_0 
\end{equation}
because nRule (4) will not allow $\underline{D}_w(t)$ to pass probability current to the other components.  Component
$\underline{D}_w(t)$ is `ready' because it is a discontinuous and noncyclic quantum jump away from $\psi(t)D_0B_0$ and it is complete
because it includes the (not shown) entangled environment.  Again, the time dependence of $\underline{D}_w(t)$ does not
represent a dynamical evolution beyond the changes given to it by the first component.  It does not evolve dynamically
on its own.  And again, the sub-0 on $B_0$ indicates an awareness of the ground state $D_0$ of the detector.  Since the second brain
state in Eq.\ 5 is the same as the first, there is only one brain state $B_0$ in this superposition.  A cat-like ambiguity is
thereby avoided.  Equation 5 now \emph{replaces} Eq.\ 4.   

	Equation 5 is the state of the system before there is a stochastic hit that produces a state reduction.  If there is a capture,
then there will be a stochastic hit on the second component of Eq.\ 5 at a time $t_{sc}$. This will reduce the first component to
zero according to nRule (3), and convert the ready state in the second component to a realized state.
\begin{displaymath}
\Phi(t = t_{sc} > t_0) = D_w(t)B_0
\end{displaymath}
The observer is still conscious of the detector's ground state in this equation because the capture has only affected the window end
of the detector.  But after $t_{sc}$, a continuous evolution will produce
\begin{equation}
\Phi(t \ge t_{sc} > t_0) = D_w(t)B_0 \rightarrow D_{m}(t)B_0 \rightarrow D_d(t)B_1
\end{equation}
Since this equation represents a \emph{single component} that evolves in time as shown, there is no time at which both $B_0$ and
$B_1$ appear simultaneously.  There is therefore no cat-like ambiguity in this equation.

\vspace{.4cm}

	Standard quantum mechanics  gives us Eq.\ 4 by the same logic that it gives us Schr\"{o}dinger's cat and
Everett's many worlds.  \mbox{Equation 4} (top or bottom row) is a single equation that simultaneously presents two different
conscious brain states, resulting in an unacceptable ambiguity.  But with the nRules, each Schr\"{o}dinger solution is
separately grounded by its own stochastically selected boundary, allowing the rules to correctly and unambiguously predict the
continuous experience of the observer in two stages.  This is accomplished by replacing `one' equation in \mbox{Eq.\ 4} with `two'
equations in Eqs.\ 5 and 6.   Equation 5 describes the state of the system \emph{before} capture, and Eq.\ 6 describes the state of
the system \emph{after} capture.  Before and after are two \emph{different} solutions to Schr\"{o}dinger's equation, specified by
different boundary conditions.  Remember we said that the stochastic trigger selects the new boundary that applies to the reduced
state.  So it is the stochastic event that separates the two solutions \mbox{-- defining} before and after.  

		If there is no stochastic hit on the second component in Eq.\ 5 it will become a phantom component.  The new
system is then just the first component of that equation.  This corresponds to the observer continuing to see the ground state
detector $D_0$, as he should in this 
case\footnote{Had the reasoning of Footnote 5 been applied to Eq.\ 4, the component $D_1(t)B_1$
would be a superposition of the continuum of launch possibilities.  It would include a superposition of all the brain states $B_0$
that existed before the signal could have traveled through the detector to reach the brain, plus all the brain states $B_1$ that
were reached after that time.  This would have produced a massive cat-like paradox prior to a stochastic hit or state reduction of
any kind.  However, the nRules also avoid this difficulty because they produce only \emph{one} launch component
$D_1(t_{sc})B_1$.}.

\section*{A Terminal Observation}
	An observer who is inside a system must be able to confirm the validity of the Born rule that is normally applied from the
outside.  To show this, suppose our observer is not aware of the detector during the interaction with the particle, but
he looks at the detector after  a time $t_f$ when the primary interaction is
complete. Assume initial normalization equal to 1.0. During that interaction we have 
\begin{equation}
\Phi(t_f > t \ge t_0) = [\psi(t)d_0 + \underline{d}_w(t)]\otimes X
\end{equation}
where $X$ is the unknown state of the observer prior to the physiological 
interaction. 

Assume there has \emph{not} been a capture.  Then after the interaction is complete and before the observer looks at the detector
we have
\begin{displaymath}
\Phi(t \ge t_f > t_0) = [\psi(t)d_0 + \underline{d}_w(t_f)]\otimes X
\end{displaymath}
where there is no longer a probability current flow inside the bracket. The second component in the bracket is therefore a
phantom.  There is no current flowing into it, and none can flow out of it because of nRule (4).  So the equation is
essentially   
\begin{displaymath}
\Phi(t \ge t_f > t_0) = \psi(t)d_0\otimes X
\end{displaymath}
When the observer finally observes the detector at $t_{ob}$ he will get
\begin{displaymath}
\Phi(t \ge t_{ob} > t_f > t_0) = \psi(t)d_0\otimes X \rightarrow \psi(t)D_0B_0 
\end{displaymath}
where the physiological process (represented by the arrow) carries $\otimes X$ into $B_0$ and $d_0$ into $D_0$ by a continuous
classical progression leading from independence to entanglement.   This corresponds to the observer coming on board to witness
the detector in its ground state as he should when is no capture.  The probability of this happening (according to the sRules) is
equal to the square modulus of $\psi(t)d_0\otimes X$ in Eq.\ 7.

If the particle \emph{is} captured during the primary interaction, there will be a stochastic hit on the second component inside
the bracket of Eq.\ 7 at some time $t_{sc} < t_f$.  This results in a capture given by
\begin{displaymath}
\Phi(t_f > t = t_{sc} > t_0) = d_w(t)\otimes X
\end{displaymath}
after which 
\begin{displaymath}
\Phi(t_f > t \ge  t_{sc} > t_0) = [d_w(t) \rightarrow d_m(t) \rightarrow d_d(t)]\otimes X 
\end{displaymath}
\begin{displaymath}
\Phi(t \ge t_f >   t_{sc} > t_0) =  d_1(t)\otimes X
\end{displaymath}
as a result of the classical progression inside the detector.  When the observer does become aware of the detector at $t_{ob} > t_f$
we finally get
\begin{displaymath}
\Phi(t \ge t_{ob} > t_f > t_{sc} > t_0) = d_1\otimes X \rightarrow D_1B_1
\end{displaymath}
So the observer comes on board to witness the detector in its capture state with a probability (according to the sRules) equal to
the square modulus of
$\underline{d}_w(t_f)\otimes X$ in Eq.\ 7.  The nRules therefore confirm the Born rule, in this case as a theorem.

\section*{An Intermediate Case}
		In Eq.\ 5 the observer is assumed to interact with the detector from the beginning.  Suppose that the incoming particle results
from a long half-life decay, and that the observer's physiological involvement only \emph{begins}  in the middle of the
primary interaction.  Before that time we will have
\begin{displaymath}
\Phi(t \ge t_0) = [\psi(t)d_0 + \underline{d}_w(t)]\otimes X
\end{displaymath}
where again $X$ is the unknown brain state of the observer prior to the physiological interaction.  Primary probability current here
flows between the detector components inside the bracket.  Let the physiological interaction begin at a time $t_{look}$ after
$\underline{d}_w(t)$ has gained some amplitude, and suppose it lasts for a period of time equal to $\pi$.   Then
\begin{eqnarray}
&\Phi(t=t_{look} > t_0) = \psi(t)d_0\otimes X + \underline{d}_w(t) \otimes X&  \\
&\hspace{3.2cm}\downarrow\hspace{1.6cm}\downarrow&\nonumber\\
&\Phi(t=t_{look} + \pi > t_{look} > t_0) = \psi(t)D_0B_0 + \underline{D}_w(t) B_0& \nonumber
\end{eqnarray}
where the arrows mean that the evolution from the first to the second row is classical and continuous. That evolution carries
$\otimes X$ into $B_0$ and $d_0$ into $D_0$ by a process that leads from independence to entanglement.  So while the second
component in each row gains square modulus because of the primary interaction (plus signs), the observer is simultaneously coming on
board by a continuous process (arrows).  

The second component at each moment of time during the physiological interaction in Eq.\ 8 is the launch state of the new solution
of the Schr\"{o}dinger equation -- if there is a stochastic hit at that moment.  This component establishes the boundary conditions
of any  newly launched solution.   

After the time $t_{ob} =  t_{look} + \pi$ we can write Eq.\ 8 as 
\begin{equation}
\Phi(t \ge t_{ob} > t_0) = \psi(t)D_0B_0 + \underline{D}_w(t)B_0
\end{equation}
This equation is identical with Eq.\ 5; so from this moment on, it is as though the observer has been on board from the beginning.  

  	If there is a subsequent capture at a time $t_{sc}$, this will become like Eq.\ 6.
\begin{equation}
\Phi(t \ge t_{sc} > t_{ob} > t_0) = D_w(t)B_0 \rightarrow  D_m(t)B_0 \rightarrow D_d(t)B_0
\end{equation}

	If a stochastic hit occurs between $t_{look}$ and $t_{ob}$, then the ready component at that moment (the second component in Eq.\
8) will be chosen and made a realized state.  It will then proceed classically and continuously to $D_dB_1$ as in Eq.\ 10.

\section*{A Second Observer}
If a second observer is standing by while the first observer interacts with the detector during the primary interaction, the state
of the system will be
\begin{displaymath}
\Phi(t \ge  t_0) = [\psi(t)D_0B_0 + \underline{D}_w(t)B_0]\otimes X
\end{displaymath}
where $X$ is an unknown state of the second observer prior to his interacting with the system.  The detector $D$ here includes the
low-level physiology of the first observer.  A further expansion of the detector will include the second observer's low-level
physiology when he comes on board.  The detector will then split into two parallel paths, one connecting to the first observer and
the other  connecting to the second observer.    When a product of brain states appears in the form $BB$ or $B\otimes X$, the
first term will refer to the first observer and the second to the second observer.  
	
The result of the second observer looking at the detector will be the same as that found for the first observer in the previous
section, except now the first observer will be present in each case.  In particular, the equations similar to \mbox{Eqs.\ 8, 9, and
10} are now 
\begin{eqnarray}
&\Phi(t=t_{look} + t) = \psi(t)d_0\otimes X + \underline{d}_w(t) \otimes X \rightarrow& \nonumber\\
 &\Phi(t = t_{look} + \pi > t_{look} > t_0) = [\psi(t)D_0B_0B_0 + \underline{D}_w(t)B_0B_0]&  \nonumber\\
&\Phi(t \ge t_{ob}  > t_0) = \psi(t)D_0B_0B_0 + \underline{D}_w(t)B_0B_0&  \nonumber \\
&\Phi(t \ge t_{sc} > t_{ob}  > t_0) = D_w(t)B_0B_0 \rightarrow D_m(t)B_0B_0 \rightarrow D_d(t)B_1B_1&  \nonumber 
\end{eqnarray}
These will all yield the same result for the new observer as they did for the old observer.  In no case will the nRules produce a
result like $B_1B_2$ or $B_2B_1$.

\vspace{.4cm}

Up to this point we have seen how the nRules go about including observers inside a system in an ontological model.  These rules
describe when and how the observer becomes conscious of measuring instruments, and replicate common empirical experience in these
situations. The nRules are also successfully applied in another paper \cite{RM5}  where two versions of the Schr\"{o}dinger cat
experiment are examined.  In the first version a conscious cat is made unconscious by a stochastically initiated process; and in the
second version an unconscious cat is made conscious by a stochastically initiated process. 

In the following sections we turn attention to another problem -- the requirement that macroscopic states must appear in
their normal sequence.   This sequencing chore represents a major application of nRule (4) that is best illustrated in the case of a
macroscopic counter.

\section*{A Counter}

If a beta counter that is exposed to a radioactive source is turned on at time $t_0$, its state function will be given by
\begin{displaymath}
\Phi(t \ge   t_0) = C_0(t) + \underline{C}_1(t)\
\end{displaymath}
where $C_0$ is a counter that reads zero counts, $C_1$ reads one count, and $C_2$ (not shown) reads two counts, etc.  The second
component $\underline{C}_1(t)$ is zero at $t_0$ and increases in time.  The underline indicates that it is a ready state as
required by nRule (2).  $C_2$ and higher states do not appear in this equation because \mbox{nRule (4)} forbids current to leave
$\underline{C}_1(t)$.  Ignore the time required for the capture effects to go from the window to the display end of the counter.  

The $0^{th}$ and the $1^{st}$ components are the only ones that are initially active, where the current flow is $J_{01}$
from the $0^{th}$ to the $1^{st}$ component.  The resulting distribution at some time $t$ before $t_{sc}$ is shown in Fig.\ 3,
where $t_{sc}$ is the time of a stochastic hit on the second component.  

\begin{figure}[h]
\centering
\includegraphics[scale=0.7]{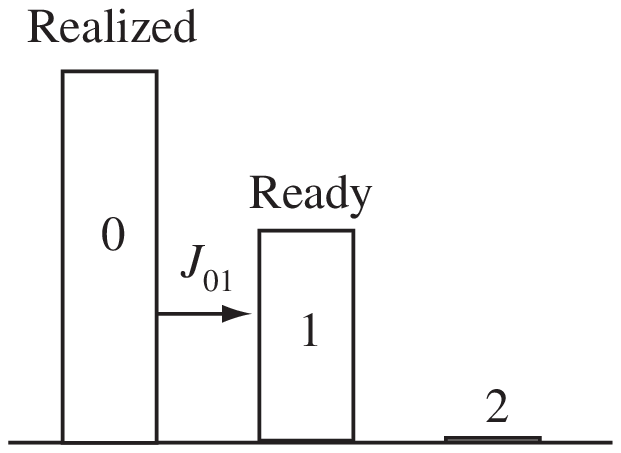}
\center{Figure 3: Square moduli before the first stochastic hit}
\end{figure}

This means that the $1^{st}$ component \emph{will} be chosen because all of the current from the (say,
normalized) $0^{th}$ component will pore into the $1^{st}$ component making $\int J_{01}dt = 1.0$.  Following the stochastic hit
on the $1^{st}$ component, there will be a collapse to that component because of \mbox{nRule (3)}.  The first two dial readings
will therefore be sequential, going from 0 to 1 without skipping a step such as going directly from 0 to 2.  It is nRule (4) that
enforces the no-skip behavior of a counter; for without it, any component in the superposition might be chosen as a result of
probability current flowing into it.    It is empirically mandated that these states should always follow in sequence without
skipping a step. 

\begin{figure}[h]
\centering
\includegraphics[scale=1.2]{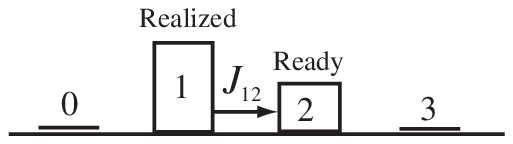}
\center{Figure 4: Square moduli before second hit}
\end{figure}

With the stochastic choice of the $1^{st}$ component at $t_{sc}$, the process will begin again as shown in 
Fig.\ 4.  This  leads with certainty to a stochastic choice of the $2^{nd}$ component.  That certainty is accomplished by the
wording of nRule (1) that requires current normalization at each moment of time; that is, the  current  $J_{12}$ is divided by the
total square modulus at each moment.  The total integral $\int J_{12}dt$  is less than 1.0 in  Fig.\ 4 because of the reduction that
occurred in \mbox{Fig.\ 3};  but it is restored to 1.0 when divided by the (new) total square modulus.  It is therefore certain that
the $2^{nd}$ component will be chosen.

And finally, with the choice of the $2^{nd}$ component, the process will resume again with current $J_{23}/s$ going from the
$2^{nd}$ to the $3^{rd}$ component.  This leads with certainty to a stochastic choice of the $3^{rd}$ component.  

 If an observer watches the counter from the beginning he will be able to see it go sequentially from $C_0$ to $C_1$, to $C_2$,
etc., for he is himself a continuous part of the system.  He does not have to `peek' intermittently like an
epistemological observer who is not part of the system as has been said.  Although the empirical experience of the ontological
(nRule) observer is different from the epistemological (sRule) observer, there is no contradiction between the two.  That's because
the nRules and the sRules answer two different questions about probability as previously noted.  The nRules ask: What is the
probability that the system will jump to the \emph{next} counter state in the next differentially small increment of time; and the
sRules ask: What it the probability that an outside observer will find the system in any one of its possible counter states if he
looks after some finite time $T$?

\section*{The Parallel Case} 
Now imagine parallel states in which a quantum process may go either clockwise or counterclockwise as shown in Fig. 5.  Each
component includes a macroscopic piece of laboratory apparatus $A$, where the Hamiltonian provides for a discontinuous clockwise
interaction going from the $0^{th}$ to the $r^{th}$ state, and another one going from there to the final state $f$; as well as a
comparable counterclockwise interaction from the $0^{th}$ to the $l^{th}$ state and from there to the final \mbox{state $f$}.  The
Hamiltonian does not provide a direct route from the $0^{th}$ to the final state, so the system will choose stochastically between a
clockwise and a counterclockwise route.  Ready states $\underline{A}_l$ and $\underline{A}_r$ are the \emph{eigenstates} of that
choice, and contain the boundary conditions for each separate path.

\begin{figure}[t]
\centering
\includegraphics[scale=0.8]{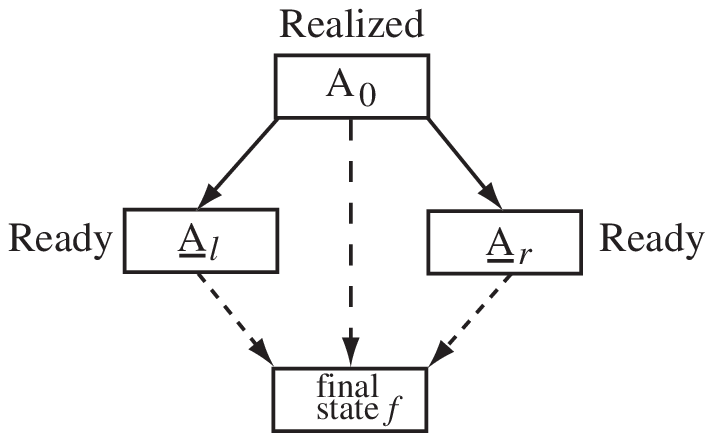}
\center{Figure 5: Possible parallel decay routes}
\end{figure}

With nRule (4) in place, probability current cannot initially flow from either of the intermediate states to the final state, for
that would require current flow from a ready state.  The dashed lines in Fig.\ 5 indicate the forbidden
transitions.  But once the state $\underline{A}_l$ (or $\underline{A}_r$) has been stochastically chosen, it will become a
realized state $A_l$ (or $A_r$) and a subsequent transition to $\underline{A}_f$ can occur that realizes $A_f$.  

The effect of nRule (4) is therefore to force this macroscopic system into a classical sequence that goes either clockwise or
counterclockwise.  Without it, the system might make a  second order transition (through one of the intermediate states) to the final
state, without the intermediate state being realized.  The observer would then see the initial state followed by the final state,
without knowing (in principle) which pathway was followed.  This is a familiar property of continuous microscopic evolution.  In
Heisenberg's famous example formalized by Feynman, a microscopic particle observed at point $a$ and later at point $b$ will travel
over a quantum mechanical superposition of all possible paths in between.  Without nRule (4), macroscopic objects facing
discontinuous and noncyclic parallel choices would do the same thing.  But that should not occur.  The fourth nRule forces this
parallel system into one or the other classical path, so it is not a quantum mechanical superposition of both paths.

\section*{A Continuous Variable}
In the above examples nRule (4) guarantees that sequential discontinuous steps in a superposition are not passed over. 
If the variable itself is classical and continuous, then continuous observation is possible without the
necessity of stochastic jumps.  In that case we do not need \mbox{nRule (4)} or any of the \mbox{nRules (1-4)}, for they do not
prevent or in any way qualify the motion.  

However, a classical variable may require a quantum mechanical jump-start.  For instance, the
mechanical device that is used to  seal the fate of Schr\"{o}dinger's cat (e.g., a falling hammer) begins its motion with a
stochastic hit.  That is, the decision to begin the motion (or not) is left to a $\beta$-decay.  In this case \mbox{nRule (4)}
forces the motion to begin at the beginning, insuring that no value of the classical variable is passed over;  so the hammer
will always fall from its \emph{initial} angle with the horizontal.   Without nRule (4), the hammer might begin its fall at some
other angle because probability current will flow into angles other than the initial one.

\section*{Microscopic Systems}
The discussion so far has been limited to experiments or procedures whose outcome is empirically known.  Our claim has been that the
nRules are chosen to work without regard to a theory as to `why' they work.  Therefore, our attention has always gone to
macroscopic situations in which the results are directly available to conscious experience.  However, if the nRules are correct we
would also want know how they apply to microscopic systems, even though the predicted results in these cases are more
speculative.  In this section we will consider the implications of the nRules in three microscopic cases.  The important question to
ask in each case is: Under what circumstances will these rules result in a state reduction of a microscopic system?

\begin{displaymath}
\mbox{\textbf{Case 1 -- spin states}}
\end{displaymath}

No state reduction will result from changing the representation.    In particular, replacing the
spin state $+z$ with the sum of states $+x$ and $-x$ will not result in either one of the $x$-states becoming a ready state.  

	This will be true for a spin $+z$ particle even if the common environment of $+x$ and $-x$ includes a magnetic field that is
continuously applied in (say) the $x$-direction.  So long as the magnetic field  is the same for both $+x$ and $-x$, the result will
be the same (i.e., neither one will become a ready state).  

More generally, let the environments $E^p(t)$ of $+x$ and $E^n(t)$ of $-x$ change continuously over time such that
\begin{equation}
\Phi(\tau > t \ge t_0) = \frac{1}{\sqrt{2}}(+z)E^p(t) + \frac{1}{\sqrt{2}}(-z)E^n(t)
\end{equation}
where $E^p(0) = E^n(0) = E$, and $E^p(\tau) = E_+,  E^n(\tau) = E_-$.  States $E_+$ and $E_-$ are the final environments of $+x$
and $-x$ at a time $\tau$ that will be either the time at which the process is complete, or  complete to some
desired extent.  It may be that the environments $E_+$ and $E_-$ are \emph{similar} to $E$ and to each other (i.e., same temperature
and pressure, same particles, same radiation field, etc.), but time will change $E_+$ and $E_-$ so they can no longer be
\emph{identical} with $E$ or with each other.  Of course, it might also be that $E_+$ and $E_-$ are not even similar \mbox{to E}.

Taking the first and last term in Eq.\ 11 we write
\begin{eqnarray}
\Phi(\tau > t \ge t_0) &=& \frac{1}{\sqrt{2}}[(+z) + (-z)]E \rightarrow \frac{1}{\sqrt{2}}[(+z)E_+ + (-z)E_-] \\
&&   \hspace{1cm}  t =t_ 0  \hspace{2.8cm} t = \tau  \nonumber
\end{eqnarray}
where the arrow indicates a continuous transition. Therefore, there will be \emph{no state reduction} in this case.  It may be
noncyclic, but this process does not lead to a collapse of the wave to either $+x$ or $-x$.   

	For example, if the magnetic field is non-homogeneous there will be a physical separation of the $+x$ and $-x$ states.  One will
move into a stronger magnetic field and the other will move into a weaker magnetic field.  Assuming that this field is continuously
(i.e., classically) applied, there will be no ready state and no state reduction.  Ready states will appear only when the $+x$ and
$-x$ components are picked up by different detectors at different locations, resulting in a \emph{detector related} discontinuity. 
According to nRule (3), a state reduction can only occur if and when that happens.

	There are many other `quantum' processes that are continuous and therefore not subject to stochastic reduction, such
as scattering, interference, diffraction, and tunneling.  Vacuum fluctuations are cyclic. 

\begin{displaymath}
\mbox{\textbf{Case 2 -- Free neutron decay}}
\end{displaymath}

A free neutron decay is written $\Phi(t) = n(t) + \underline{ep\overline{\nu}}(t)$, where the second component is zero at $t = 0$
and increases in time as probability current flows into it.  This component contains three entangled particles  making a whole 
object, where all three  are `ready' states as indicated by the underline \mbox{(see nRule 2)}. Each component  is multiplied by a
term representing the environment (not shown).  Each  is complete for this reason and also because the variables of each particle
take on all of the values that are allowed by the boundary conditions.  Following \mbox{nRule (3)} there will be a
stochastic hit on $\underline{ep\overline{\nu}(t)}$ at some time $t_{sc}$, reducing the system to the realized correlated states
$ep\overline{\nu}(t_{sc})$.  

Specific values of, say, the electron's momentum are not stochastically chosen by this reduction because all possible values of
momentum are included in $ep\overline{\nu}(t_{sc})$ and its subsequent evolution.  For the electron's momentum to be determined in
a specific direction away from the decay site, a detector in that direction must be activated.  That will require another stochastic
hit on the component that includes that detector.   

This case provides a good example of how  $\underline{ep\overline{\nu}}(t)$ is a function time beyond its increase in square
modulus.  Assume that the neutron moves across the laboratory in a wave packet of finite width.  At each moment the ready component
$\underline{ep\overline{\nu}}(t)$ will ride with the packet, having the same size, shape, and group velocity.  It is the launch
component that contains the boundary conditions of the next solution of Schr\"{o}dinger's equation -- the solution that appears when
$\underline{ep\overline{\nu}}(t)$ is stochastically chosen at $t_{sc}$.  Before this `collapse', $\underline{ep\overline{\nu}}(t)$
is time dependent because it increases in square modulus \emph{and} because it follows the motion of the neutron; however, nRule (4)
insures that it will not evolve dynamically beyond itself until it becomes a realized component at the time of  stochastic choice. 
The neutron
$n(t)$ will then disappear and the separate particles $e(t)p(t)\overline{\nu}(t)$ will spread out on their own, still
correlated in conserved quantities.   

\begin{displaymath}
\mbox{\textbf{Case 3 -- Atomic absorption and emission}}
\end{displaymath}

If an atom is raised to an excited state by a passing photons, the absorption part of this interaction is cyclic because the atom
might fall back to its ground state by a stimulated emission.  However, the excited state atom might also emit `another' photon by
simultaneous emission and that is noncyclic.  That could lead to a stochastic hit and collapse of the wave to the newly formed decay
state.  The  pulse of $N$ photons is represented by $\gamma_N(t)$ in the first component of Eq.\ 13, and the spontaneously emitted
photon is $\underline{\gamma}(t)$ in the third component. 

\begin{equation} 
\Phi(t \ge t_0) = \gamma_N(t)A_0(t) \leftrightarrow \gamma_{N-1}(t)A_1(t)+ \gamma_{N-1}(t)\underline{\gamma}(t)\underline{A}_0(t)
\end{equation}
The continuous (cyclic) oscillation is represented by the reversible arrow $\leftrightarrow$ between the first and second
components.

If the incoming photons have a small cross section with the atom, the second component in Eq.\ 13 will not fully discharge into the
third component.  In that case the latter will become a phantom that can be subsequently ignored.  There will be no collapse.  The
second component will then dampen out and the first component, modified by the encounter, will be the only survivor.   

Otherwise, there will be a stochastic hit on the third component at some time $t_{sc}$, resulting in 
\begin{displaymath}
\Phi(t \ge t_{sc} < t_0) = \gamma_{N-1}(t)\gamma(t)A_0(t)
\end{displaymath}  
The newly emitted photon $\gamma(t)$ will evolve until it has a final pulse width $\Delta T_f$ that is
associated with its final spread of energy $\Delta E$.  There is no necessary relationship between $\Delta T_f$ and the half-life
$T_{1/2}$ of the decay from the second to the third component in Eq.\ 13.  There might be any number of influences affecting
$T_{1/2}$, including the number of photons in the stimulating pulse $\gamma_N(t)$.  If there are no incoming photons, and if the
excited state $A_1(t)$ is realized by a quantum jump from a higher energy level at some stochastically chosen moment $t_{scc}$,
then 
\begin{equation} 
\Phi(t \ge t_{scc}) = A_1(t)+ \underline{\gamma}(t)\underline{A}_0(t)
\end{equation}
where the second component is zero at $t_{scc}$ and increases in time.  The state $A_1(t)$  decays \emph{only} into the
second component in this equation; whereas in \mbox{Eq.\ 13} it shares time with the initial state
$\gamma_N(t)A_0(t)$.  In Eq. 13 the  average current flow into $\underline{\gamma}(t)\underline{A}_0(t)$ is therefore decreased,
so the half-life $T_{1/2}$ of the spontaneous decay is increased.

In contrast, $\Delta T_f$ is determined only by the time dependent solutions of the excited state $A_1(t)$.  So $\Delta T_f$ is
not causally connected to $T_{1/2}$.  The value of $\Delta T_f$ is transmitted from $A_1(t)$ to the launch component
$\underline{\gamma} (t)\underline{A}_0(t)$ at each moment of time, setting the boundary conditions of that launch site at that
moment.  Therefore, it will not matter to the properties of the emitted photon (in particular $\Delta T_f$) if a stochastic hit
on $\underline{\gamma}(t)\underline{A}_0(t)$ in Eq.\ 13 or Eq.\ 14 aborts the decay before it is complete.

\section*{Decoherence}
Suppose that two states $A$ and $B$ that are initially in coherent Rabi oscillation and are exposed to a phase disrupting
environment.  This may be expressed by the equation
\begin{equation}
\Phi(\tau > t \ge t_0) = (A \leftrightarrow B)E(t) \rightarrow [AE_A(t) + BE_B(\tau)] 
\end{equation}
where $A$ is at a higher energy level than $B$.  The right pointing arrow indicates a continuous process.  When the environments
$E_A$ and $E_B$ are sufficiently different to be orthogonal, then states $A$ and $B$ in Eq.\ 15 will become totally decoherent. 
Statistically, this leaves a local mixture of $A$ and $B$ with no current flow between them.  The subsequent 
decay of $AE_A$ is generally much slower than decoherence, so decoherence will be essentially complete before there is a
stochastic interruption.  Evidence for this is found in low temperature experiments \cite{DV, YY} where Rabi oscillations undergo
decoherent decay without any sign of interruption due to state reduction.     

\vspace{.4cm}

The ammonia molecule is generally found in a partially decoherent state.  In a rarified atmosphere the molecule will most likely be
in its symmetric coherent form $(A\leftrightarrow B)$, where $A$ has the nitrogen atom on one side of the
hydrogen plane, and $B$ has the nitrogen atom placed symmetrically on the other side.  This is the lowest energy state available to
the molecule.  In this case the states $A$ and $B$ (Eq.\ 15) taken by themselves are equally energetic at a higher energy level.  
  
Collisions with other molecules in the environment will tend to destroy the coherence between $A$ and $B$, causing the ammonia
molecule to become decoherent to some extent.  This decoherence can be reversed by decreasing the pressure.  Since an ammonia
molecule wants to fall into its lowest energy level, it will tend to return to the symmetric  state when outside pressure
is reduced.  In general, equilibrium can be found between a given environment and some degree of decoherence.    

	The ammonia molecule  cannot assume the symmetric form $(A \leftrightarrow B)$ if the environmental collisions are too frequent
-- i.e., if the pressure exceeds about 0.5 atm.\ at room temperature \cite{JZ}.   At low pressures the molecule is a stable coherent
system, and at high pressures it is a stable decoherent system.  It seems to change from a microscopic object to a macroscopic
object as a function of its environment.  This further supports the idea that the micro/macroscopic distinction is not
fundamental.  

\vspace{.4cm}

To accommodate the main example of this paper, which is a detector that may or may not capture a
particle, a more general form is adopted.  We now use time dependent coefficients $m(t)$ and $n(t)$, where $m(t_0) = 1, n(t_0) = 0$,
and where $m(t)$ decreases in time keeping
$m(t)^2 + n(t)^2 = 1$.  These coefficients describe the progress of the primary interaction, giving  
\begin{eqnarray}
\Phi(\tau > t \ge t_0) &=& [m(t)A + n(t)\underline{B}]E \rightarrow [m(t)AE_A(t) + n(t)\underline{B}E_B(t)] \nonumber\\
&&   \hspace{.95cm}  t = t_0  \hspace{3.3cm} t = \tau \nonumber
\end{eqnarray} 
where the transitions \emph{inside} the brackets are now discontinuous and noncyclc, making $\underline{B}$ a ready state as
required by nRule (2).  This can be written
\begin{eqnarray}
\Phi(t = t_0) &=& AE \hspace{.5cm}\mbox{(intial  component)}\hspace{1.8cm} m(t_0) = 1 \\
\Phi(\tau \ge t \ge t_0) &=&  AE \rightarrow [m(t)AE^m(t) + n(t)\underline{B}E^n(t)]\hspace{.4cm}  m(t) \rightarrow 0 
\nonumber
\end{eqnarray} 
Since state $\underline{B}$ in this equation is a `ready' state and  the  primary current flows from $A$ to
$\underline{B}$ inside the  square bracket, $\underline{B}$ is a candidate for state reduction according to nRule (3).  

	Equation 16 applied to our example in Eq.\ 3 with $m(t)A = \psi(t)d_0$ and $n(t)\underline{B} = \underline{d}_w(t)$ says that 
initially coherent states $\psi(t)d_0$ and $\underline{d}_w$ rapidly become decoherent.  Because of the macroscopic nature of the
detector, decoherence at time $\tau$ may be assumed to be complete, and the decay time from $t_0$ to $\tau$ extremely short lived. 
The  time for any newly created pair of macroscopic objects to approach full decoherence is so brief that it is not measurable in
practice.  Still, we see that decoherence does not happen immediately when a new macroscopic state is  created.

\section*{Grounding the Schr\"{o}dinger Solutions}
	Standard quantum mechanics is not completely grounded because it does not recognize all of the boundary conditions (beyond 
the initial conditions) that are stochastically chosen during the lifetime of the system.  With either one of the proposed
rule-sets, every stochastic hit sets a new boundary (i.e., the chosen eigenvalue) for a new solution of Schr\"{o}dinger's equation. 
On the other hand, traditional quantum theory accumulates all the possible solutions as though they were all simultaneously valid;
and as a result, this model encourages bizarre speculations such as the many-world interpretation of Everett or the cat paradox of
Schr\"{o}dinger.  With the proposed rule-sets, these empirical distortions disappear.  It is because standard quantum mechanics
fails to respond to the system's ongoing state reductions that these fanciful excursions seem plausible.

\section*{Limitation of the Born Rule}
Using the Born rule in standard theory, the observer can only record an observation at a given instant of time, and he must do so
consistently over an ensemble of observations.  He cannot himself become part of the system for any finite period of
time.  When discussing the Zeno effect it is said that continuous observation can be simulated by rapidly increasing the number of
instantaneous observations; but of course, that is not really  continuous.    

	 On the other hand, the observer in an ontological model can \emph{only} be continuously involved with the observed system.  Once
he is on board and fully conscious of a system, the observer can certainly try to remove himself ``immediately".  However, that
effort is not likely to result in a truly instantaneous conscious observation.    So the epistemological observer claims to make
instantaneous observations but cannot make continuous ones; and the ontological observer makes continuous observations but cannot
(in practice) make instantaneous ones.  Evidently the Born rule would require the ontological observer to do something that cannot
be realistically done.  Epistemologically we can ignore this difficulty, but a consistent ontology should not match a continuous
physical process with continuous observation by using a discontinuous rule of correspondence.  Therefore, no ontological model
should  make fundamental use of the Born interpretation that places unrealistic demands on an observer.

\section*{Status of the Rules}
No attempt has been made to relate conscious brain states to particular neurological configurations.  The nRules are an empirically
discovered set of macro relationships that exist on another level than microphysiology, and there is no need to connect these two
domains.  These rules preside over physiological detail in the same way that thermodynamics presides over molecular detail.  It is
desirable to eventually connect these domains as thermodynamics is now connected to molecular motion; and hopefully, this is what a
covering theory will do.  But for the present we are left to investigate the rules by themselves without the benefit of a wider
theoretical understanding of state reduction or of conscious systems.  There are two rule-sets of this kind giving us two different
quantum friendly ontologies -- the nRules of this paper and the oRules of Refs. 1-3.  
  
\hspace{.4cm}

The question is, which if either of these two rule-sets is correct (or most correct)?  Without the availability of a wider
theoretical structure or a discriminating observation, there is no way to tell.  Reduction theories that are currently being
considered may accommodate a conscious observer, but none are fully accepted.  So the search goes on for an extension of quantum
mechanics that is sufficiently comprehensive to cover the collapse associated with an individual measurement.  I expect that
any such theory will support one of the ontological rule-sets, so these rules might server as a guide for the construction of a
\mbox{wider theory}.

\pagebreak

\end{document}